\documentclass[aps,prb,twocolumn,floatfix,amsmath,amssymb]{revtex4}

\usepackage{lmodern}
\usepackage{graphicx}
\usepackage{bm}
\usepackage{color}
\usepackage[percent]{overpic}
\usepackage{pdfsync}
\usepackage[breaklinks=true,colorlinks,citecolor=blue,linkcolor=blue,urlcolor=blue]{hyperref}
\usepackage[normalem]{ulem}

\begin{document}

\title{Theory of the plasma-wave photoresponse of a gated graphene sheet}

\author{Andrea Tomadin}
\email{andrea.tomadin@sns.it}
\affiliation{NEST, Istituto Nanoscienze-CNR and Scuola Normale Superiore, I-56126 Pisa, Italy}

\author{Marco Polini}
\affiliation{NEST, Istituto Nanoscienze-CNR and Scuola Normale Superiore, I-56126 Pisa, Italy}


\begin{abstract}
The photoresponse of graphene has recently received considerable attention.
The main mechanisms yielding a finite dc response to an oscillating radiation field which have been investigated include responses of photovoltaic, photo-thermoelectric, and bolometric origin.
In this Article we present a fully analytical theory of a photoresponse mechanism which is based on the excitation of plasma waves in a gated graphene sheet.
By employing the theory of relativistic hydrodynamics, we demonstrate that plasma-wave photodetection is substantially influenced by the massless Dirac fermion character of carriers in graphene and that the efficiency of photodetection can be improved with respect to that of ordinary parabolic-band electron fluids in semiconductor heterostructures.
\end{abstract}

\maketitle

\section{Introduction}

The potential of graphene~\cite{geim_naturemater_2007,castroneto_rmp_2009,peres_rmp_2010,dassarma_rmp_2011,kotov_rmp_2012,Katsnelsonbook, bonaccorso_matertoday_2012}---a two-dimensional (2D) crystal of Carbon atoms tightly packed in a honeycomb lattice---in optoelectronics, photonics, and plasmonics is attracting a truly considerable attention~\cite{bonaccorso_naturephoton_2010,koppens_nanolett_2011,grigorenko_naturephoton_2012}.

The particle-hole-symmetric spectrum of massless Dirac fermions~\cite{geim_naturemater_2007,castroneto_rmp_2009,peres_rmp_2010,dassarma_rmp_2011,kotov_rmp_2012,Katsnelsonbook} in graphene and its structureless optical conductivity~\cite{universal} are particularly suitable~\cite{bonaccorso_naturephoton_2010} to realize detectors of radiation in a wide range of photon energies from visible to terahertz (THz) frequencies.
As a consequence, the photoresponse of a graphene sheet has been the subject of truly intense investigations~\cite{xia_naturenano_2009, mueller_prb_2009,mueller_naturephoton_2010,xu_nanolett_2010,peters_apl_2010,echtermeyer_naturecommun_2011,lemme_nanolett_2011,song_nanolett_2011,gabor_science_2011,rao_acsnano_2011,konstantatos_naturenano_2012,vora_apl_2012,yan_naturenano_2012,vicarelli_naturemat_2012,freitag_naturephoton_2013,freitag_nanolett_2013}.
Three main photoresponse mechanisms have been identified in the literature to date.
(i) A finite dc response to an oscillating radiation field can simply stem from the ordinary {\it photovoltaic} effect.
Photons impinging on the sample excite electron-hole pairs and their basic constituents (the electron and the hole) are separated by the electric field across a p-n junction.
(ii) It has been recognized that hot-carrier-assisted transport plays an important role in graphene~\cite{xu_nanolett_2010}.
Due to the large optical phonon energy scale in this material, hot carriers created by the radiation field can remain at a temperature higher than that of the lattice for several tens of ${\rm ps}$.
Equilibration with the lattice occurs indeed mainly because of scattering between carriers and acoustic phonons~\cite{bistritzer_prl_2009,tse_prb_2009}.
These processes take place on a ${\rm ns}$ timescale, although they can experience a speed-up in the case of disorder-assisted collisions~\cite{song_prl_2012, graham_naturephys_2012, betz_naturephys_2012}.
Hot carriers can therefore significantly alter the photoresponse of a graphene sheet~\cite{xu_nanolett_2010,song_nanolett_2011,gabor_science_2011} by virtue of the {\it photo-thermoelectric} effect.
Carrier multiplication~\cite{brida_naturecommun_2013,tomadin_arxiv_2013} can greatly enhance the performance of graphene photodetectors operating on the basis of photovoltaic and photo-thermoelectric effects. 
(iii) Finally, {\it bolometric} effects can play an important role in the photoresponse of a graphene sheet.
In this case, the radiation impinging on the sample produces heating, which, in turn, affects the magnitude of the resistance.

\begin{figure}
\includegraphics[width=\linewidth]{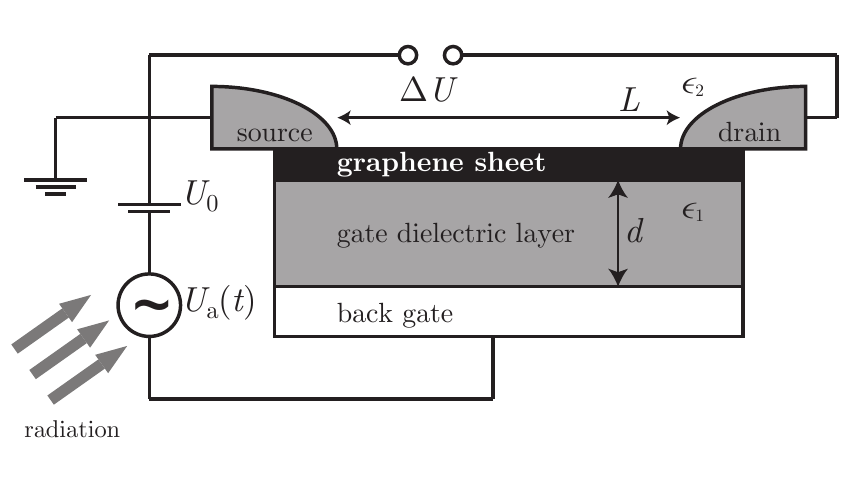}
\caption{\label{fig:one}
Schematics of a Dyakonov-Shur single-layer-graphene field-effect transistor for photodetection.
An oscillating radiation field coupled to a graphene field-effect transistor creates an oscillating potential difference $U(t) = U_0 + U_{\rm a}\cos(\Omega t)$ between the back gate and source. 
The graphene flake is deposited on a substrate with dielectric constant $\epsilon_1$ and is separated by a distance $d$ from the back gate.
The dielectric constant of the material above the graphene sheet is $\epsilon_2$.
The separation between source and drain is $L$.
Because of the intrinsic nonlinear response of the two-dimensional electron fluid in the field-effect transistor channel, a finite dc potential difference $\Delta U$, which is proportional to the power of the incident radiation (i.e.~$\propto U^2_{\rm a}$), is measured between the source and drain at {\it zero} source-drain bias. }
\end{figure}

In a series of pioneering papers~\cite{dyakonov_prl_1993, dyakonov_prb_1995, dyakonov_ieee_1996a, dyakonov_ieee_1996b}, which appeared in the mid nineties, Dyakonov and Shur (DS) proposed a very elegant mechanism that yields a finite dc response to an oscillating radiation field.
The DS photodetection mechanism is based on the fact that a field-effect transistor (FET) hosting a 2D electron gas (2DEG) acts as a cavity for plasma waves.
When these are weakly damped, i.e.~when a plasma wave launched at the source can reach the drain in a time shorter than the momentum relaxation time $\tau$, the detection of radiation exploits constructive interference of the plasma waves in the cavity, which results in a resonantly enhanced response.
This is the so-called {\it resonant regime} of plasma-wave photodetection.
Plasma waves propagating in the FET channel cannot be simply identified with the well-known ``plasmons'' of a 2DEG~\cite{Giuliani_and_Vignale}. 
Indeed, plasma waves are collective oscillations that occur in a {\it gated} 2DEG, whereby the long-range tail of the Coulomb interaction among electrons is screened by the presence of a metal gate.
Plasma waves with wave vector $k = |{\bm k}|\lesssim 1/d$ (where $d$ is the distance between the 2DEG and the gate) have a gapless dispersion relation $\omega = s k$ (at zero temperature and neglecting friction and viscosity~\cite{svintsov_jap_2012}) which resembles that of sound waves in ordinary gases and liquids.
On the contrary, plasmons in a 2DEG with long-range Coulomb interactions have a dispersion relation which is proportional to $\sqrt{k}$ in the long-wavelength limit~\cite{Giuliani_and_Vignale}.

DS showed~\cite{dyakonov_ieee_1996a} that the photovoltage response of the 2DEG in a FET, i.e.~the electric potential difference between drain and source, contains a dc component even if the incoming field is ac, and thus provides rectification of the signal.
In the resonant regime, the dc photoresponse is characterized by peaks at the odd multiples of the lowest plasma-wave frequency.
This rectification mechanism is of purely dynamical origin and is not related to other rectification mechanisms (occurring, for the example, at the contacts) which could also be present in a real device.
Note that rectification of the signal is necessary to detect incoming radiation that exceeds the typical cutoff frequencies of circuit elements.
The DS mechanism is therefore particularly useful to detect THz radiation.

The DS mechanism relies on two facts.
(i) The reflection symmetry corresponding to the exchange of source with drain in the FET channel (see Fig.~\ref{fig:one}) is broken by the DS boundary conditions.
These boundary conditions are unusual because DS fixed the value of the current at the drain and the value of the potential at the source (instead of operating the device by fixing the current or the potential both at the source and at the drain, as is more customary).
(ii) The fact that the photovoltage averaged over a cycle of the oscillating radiation field is finite ultimately stems from the nonlinearity of the continuity equation and can be viewed as a result of the gate modulating {\it both} the electron density and the drift velocity in the channel.

A substantial amount of experimental work has been carried out on DS photodetection in 2DEGs in ordinary semiconductor heterojunctions: for recent reviews see e.g.~Refs.~\onlinecite{knap_jimtw_2009,knap_nanotech_2013}.
Recent experimental work~\cite{vicarelli_naturemat_2012} has demonstrated that the DS photodetection mechanism is active also in the case of the 2D massless Dirac fermion (MDF) fluid in a graphene FET.
Vicarelli {\it et al.}~\cite{vicarelli_naturemat_2012} have demonstrated room-temperature THz detectors based on antenna-coupled graphene FETs, which exploit the DS mechanism but display also contributions of photovoltaic and photo-thermoelectric origin.
The plasma waves excited by THz radiation in Ref.~\onlinecite{vicarelli_naturemat_2012} are overdamped and the fabricated THz detectors, although enabling large area, fast imaging of macroscopic samples, do not {\it yet} operate in the aforementioned resonant regime. 

In this Article we present a theory of DS plasma-wave photodetection in a graphene FET in the resonant regime.
We take into  account fundamental differences  between ordinary parabolic-band 2DEGs~\cite{dyakonov_prl_1993,dyakonov_prb_1995,dyakonov_ieee_1996a, dyakonov_ieee_1996b} and  2D MDF  fluids.
By employing the theory of relativistic hydrodynamics, we demonstrate in a fully analytical fashion that nonlinearities of purely relativistic origin substantially influence the response of graphene-based plasma-wave photodetectors.

This Article is organized as follows.
In Sec.~\ref{sect:model} we present a hydrodynamic theory of transport for MDFs in a gated graphene sheet.
In Sec.~\ref{sect:instability} we discuss the plasma-wave instability which arises when a graphene FET is subject to a small dc current bias. 
In Sec.~\ref{sect:photoresponse} we use the hydrodynamic theory outlined in Sec.~\ref{sect:model} to calculate the photoresponse of a graphene FET to an oscillating electromagnetic field when DS boundary conditions are applied.
Finally, in Sec.~\ref{sect:conclusions} we summarize our main findings and draw our main conclusions.

\section{Hydrodynamic theory for massless Dirac fermions in a gated graphene sheet}
\label{sect:model}

The theory of hydrodynamics~\cite{Landau06} can be applied to describe transport in a 2D electron fluid when electron-electron (e-e) collisions take place on a time scale $\tau_{\rm ee}$ which is much shorter than the typical time scale of the evolution of macroscopic variables (density, current, and energy).
At the same time, collisions with impurities and phonons, which spoil the conservation of momentum and energy, must be assumed to occur much less frequently.

In the case of present interest, we require $\omega_{\rm P} \tau_{\rm ee} \ll 1$, where $\omega_{\rm P}$ is the lowest frequency of a plasma wave in the FET.
This requirement implies that plasma waves are collective modes which can be excited in the ``collisional'' regime, when e-e interactions dominate the dynamics.
On the contrary, plasmons are collective modes which can be excited in the ``collisionless'' regime, because they can be understood, in the first instance and at long wavelengths, as the sloshing mode of the center of mass of the electron liquid.

To estimate the order of magnitude of $\omega_{\rm P} \tau_{\rm ee}$ for plasma waves, we use the following result~\cite{svintsov_jap_2012,principi_ssc_2011} for the plasma-wave group velocity
\begin{equation}
s  = v_{\rm F} \sqrt{N_{\rm f} \alpha_{\rm ee} d k_{\rm F}}~,
\end{equation}
which will be re-derived below---see Eq.~(\ref{eq:plasmaspeed}).
Here $v_{\rm F} \sim 1~{\rm nm}/{\rm fs}$ is the Fermi velocity of MDFs in graphene, $d$ is the distance between graphene and the gate (see Fig.~\ref{fig:one}), $k_{\rm F} = \sqrt{4\pi {\bar n}/N_{\rm f}}$ is the Fermi momentum corresponding to an average carrier (electron or hole) concentration ${\bar n}$, and $N_{\rm f} = 4$ is the number of fermion flavors in graphene (due to spin and valley degrees of freedom).
Finally, $\alpha_{\rm ee}$ is a dimensionless parameter that controls the strength of e-e interactions~\cite{kotov_rmp_2012}.
A microscopic definition of $\alpha_{\rm ee}$ will be given below after Eq.~(\ref{eq:plasmaspeed}).

We take $\alpha_{\rm ee} \sim 1$, $d \sim 100~{\rm nm}$, $k_{\rm F} \sim 0.17~{\rm nm}^{-1}$ (corresponding to a typical doping ${\bar n}\sim 10^{12}~{\rm cm}^{-2}$), and we find $s \sim 8.2~{\rm nm} / {\rm fs}$.
We notice that $s \gg v_{\rm F}$ with these parameters.
The largest wave vector $k$ of a plasma-wave resonance supported by the electron liquid in the FET channel is $\pi/(2 L)$, which corresponds to a standing wave with a node at the source and an antinode at the drain.
Taking $L \sim 5~\mu{\rm m}$ for a typical device, we have $\omega_{\rm P} = s k \sim 2.6 \times 10^{-3}~{\rm fs}^{-1}$ ($\hbar \omega_{\rm P} = 1.7~{\rm meV}$), corresponding to $\nu_{\rm P} = \omega_{\rm P} / (2\pi) \sim 410~{\rm GHz}$.
The transit time of a plasma wave in the device is $L / s \sim 0.61~{\rm ps}$.
A rough estimate of $\tau_{\rm ee}$ can be obtained by dividing the typical distance between two electrons ($\sim k^{-1}_{\rm F}$) by the Fermi velocity, i.e.~$\tau_{\rm ee} \sim (v_{\rm F} k_{\rm F})^{-1} \sim 5.9~{\rm fs}$.
This is in agreement with recent ultrafast two-color pump-probe measurements that have reported a time scale of the order of tens of femtoseconds to establish thermal equilibrium in single-layer graphene~\cite{brida_naturecommun_2013}.
For plasma waves in a graphene FET channel we therefore obtain $\omega_{\rm P} \tau_{\rm ee} \sim 1.5 \times 10^{-2}$, which shows that the electron system in a typical graphene FET is in the collisional regime.
On the contrary, to verify that, in the absence of a gate, plasmon excitations cannot be described within a collisional hydrodynamic theory, we consider the dispersion~\cite{grigorenko_naturephoton_2012} $\omega = v_{\rm F} \sqrt{2 \alpha_{\rm ee} k_{\rm F} k}$, evaluated at $k \sim 0.2~k_{\rm F}$.
Using the same parameters as above, we find $\omega \sim 0.11~{\rm fs}^{-1}$ ($\hbar \omega \sim 71~{\rm meV}$) and then $\omega \tau_{\rm ee} \sim 0.63$.

When the hydrodynamic assumption holds true, the system is in local equilibrium, i.e.~the electron distribution is a Fermi-Dirac distribution and does not evolve in time due to e-e collisions.
However, the parameters of the electron distribution change slowly in time and space and determine thermodynamic and hydrodynamic observables, which are not necessarily steady and evolve in time according to macroscopic conservation laws in the presence of suitable boundary conditions and slowly-varying external potentials.

Before concluding, we emphasize that for typical device lengths, the plasma-wave frequency $\nu_{\rm P}$ is in the THz regime. 
For example, changing $L$ from $1~\mu{\rm m}$ to $10~\mu{\rm m}$, the plasma-wave frequency in a graphene FET changes from $\nu_{\rm P} \sim 2.1~{\rm THz}$ to $\nu_{\rm P} \sim 210~{\rm GHz}$.
For this reason, photodetectors based on the DS mechanism are naturally useful in the context of THz light detection.

\subsection{Continuity and Euler equations}

Hydrodynamic equations~\cite{Landau06} can be derived from the Boltzmann semiclassical equation following a standard procedure~\cite{KadanoffBaym, Landau05}, which relies on the conservation of the particle number, total momentum, and total energy in e-e scattering processes.
An outline of the derivation of hydrodynamic equations for 2D MDFs in graphene is reported in the Appendix.

The continuity equation takes the familiar form
\begin{equation}\label{eq:continuity}
\partial_{t}n({\bm r},t) + \nabla_{\bm r} \cdot \lbrack n({\bm r},t) {\bm v}({\bm r},t) \rbrack  = 0~,
\end{equation}
where $n({\bm r}) = n_{\rm e}({\bm r}) + n_{\rm h}({\bm r})$ is the total carrier density and $n_{\rm e,h}({\bm r},t)$ is the electron (hole) density.
Carriers drift with the local average velocity ${\bm v}({\bm r},t)$.
The continuity equation (\ref{eq:continuity}) is identical in form to the corresponding equation for an ordinary parabolic-band 2DEG.

The Euler and energy equations, on the contrary, are dramatically different from the 2DEG case and are reported in the Appendix.
With the aim of investigating the dynamics of plasma waves in gated graphene flakes, we restrict here to the limit in which the drift velocity ${\bm v}({\bm r},t)$ is much smaller than the Fermi velocity $v_{\rm F}$.
In this case the Euler equation reads
\begin{equation}\label{eq:euler}
\begin{split}
\frac{3 P({\bm r},t)}{v_{\rm F}^{2}} \lbrace  \partial_{t} {\bm v}({\bm r},t) & + \lbrack {\bm v}({\bm r},t) \cdot \nabla_{\bm r} \rbrack {\bm v}({\bm r},t) \rbrace  = \\
& - \lbrack n_{\rm e}({\bm r},t) - n_{\rm h}({\bm r},t) \rbrack  \nabla_{\bm r} U_{\rm eff}({\bm r},t) \\
& - \nabla_{\bm r} P({\bm r},t) - \frac{{\bm v}({\bm r},t)}{v_{\rm F}^{2}} \partial_{t} P({\bm r},t)~.
\end{split}\end{equation}
Here, $P({\bm r}, t)$ is the pressure and $U_{\rm eff}({\bm r},t)$ is the effective potential which acts on electrons, including the electrostatic potential generated by the plasma wave and the fields of nearby conductors.

Note that the ``free'' parts of the Euler and energy equations [Eqs.~(\ref{eq:eulerfullrel}) and~(\ref{eq:energyfullrel}) in the Appendix, respectively] can be written in a very compact form by making use of a covariant notation, commonly employed in relativistic problems~\cite{Landau06}, where the Fermi velocity $v_{\rm F}$ plays the role of the speed of light $c$.
In the covariant notation, the Euler and energy equations correspond to the space- and time-like component of the divergence of the relativistic energy-momentum tensor~\cite{Landau06}.
This formal analogy stems from the highly-nonlinear relation [see Eq.~(\ref{eq:momentumrudin}) in the Appendix] between the average drift momentum ${\bm p}({\bm r}, t)$ and the drift velocity ${\bm v}({\bm r}, t)$, which, in turn, stems from the linear MDF dispersion of carriers in graphene~\cite{geim_naturemater_2007,castroneto_rmp_2009,peres_rmp_2010,dassarma_rmp_2011,kotov_rmp_2012,Katsnelsonbook}.
The average drift velocity ${\bm v}({\bm r},t)$ is smaller than the Fermi velocity due to the large number of e-e scattering events that take place in the hydrodynamic limit in a period of a plasma wave.

Of course, the formal analogy with relativistic fluid dynamics does not mean that the Fermi velocity is the largest velocity at which signals can causally propagate in a 2D MDF fluid.
Indeed, as shown above, the plasma-wave group velocity $s$ is typically larger than $v_{\rm F}$.
Furthermore, the Coulomb interaction between electrons and all the external potentials propagate instantaneously and do not obey relativistic causality.
The hydrodynamic theory of MDFs in graphene is therefore fundamentally different from the hydrodynamic theory of truly relativistic fluids, which is routinely employed to describe e.g.~plasmas of astrophysical interest.

We now focus our analysis on the limit $n_{\rm e}({\bm r},t) \gg n_{\rm h}({\bm r},t)$ (limit of vanishingly small hole concentration) and approximate $n({\bm r},t) \simeq n_{\rm e}({\bm r},t)$.
The local density determines the local chemical potential $\mu({\bm r},t)$ and the local Fermi energy $\varepsilon_{\rm F}({\bm r},t)$, 
\begin{equation}\label{eq:fermienergy}
\varepsilon_{\rm F}({\bm r},t) = \hbar v_{\rm F} \sqrt{4 \pi n({\bm r},t) / N_{\rm f}}~.
\end{equation}
We assume that the temperature $T$ is uniform and constant in the sample and that $k_{\rm B} T \ll \varepsilon_{\rm F}({\bm r},t)$.
The pressure [see Eq.~(\ref{eq:pressurefull}) in the Appendix] can then be approximated as
\begin{equation}\label{eq:pressure}
P({\bm r},t) \simeq \frac{1}{3} \varepsilon_{\rm F}({\bm r},t) n({\bm r},t)~.
\end{equation}

The Euler equation reduces to
\begin{equation}\label{eq:Euler_complete}
\begin{split}
\partial_{t} {\bm v}({\bm r},t) & + \lbrack {\bm v}({\bm r},t) \cdot \nabla_{\bm r} \rbrack {\bm v}({\bm r},t) = - \frac{v_{\rm F}^{2}}{\varepsilon_{\rm F}({\bm r},t)} \nabla_{\bm r} U_{\rm eff}({\bm r},t) \\
& - \frac{v_{\rm F}^{2}}{2 n({\bm r},t)} \nabla_{\bm r} n({\bm r},t)
- \frac{{\bm v}({\bm r},t)}{2 n({\bm r},t)} \partial_{t} n({\bm r},t)~.
\end{split}
\end{equation}
\subsection{Electrostatics of a graphene FET channel and the ``gradual channel'' approximation}
\label{ssec:geometry}

Let us now consider a graphene flake, which lies in the plane $z=0$ and is subject to the presence of a gate located at $z = -d$. 
The gate will be modeled as a perfect conductor.
According to Fig.~\ref{fig:one}, the dielectric constant $\epsilon(z)$ is equal to $\epsilon_{1}$ ($\epsilon_{2}$) for $-d < z <0$ ($z > 0$). 

The effective potential is given by 
\begin{equation}\label{eq:effpot}
U_{\rm eff}({\bm r},t) = -e \varphi({\bm r}',t)|_{z = 0}~,
\end{equation}
where $\varphi({\bm r}',t)$ is the three-dimensional (3D) electric potential, ${\bm r}' = ({\bm r},z)$ being a coordinate in 3D space. 

At this stage it is convenient to introduce the so-called ``gate-to-channel swing''~\cite{dyakonov_prl_1993, dyakonov_prb_1995, dyakonov_ieee_1996a, dyakonov_ieee_1996b},
\begin{equation}
U({\bm r},t) = U_{0} - \varphi({\bm r}',t)|_{z = 0}~,
\end{equation}
where $U_{0} = \varphi({\bm r}',t)\vert_{z = -d}$ is the uniform electric potential at the gate.
From this definition, $\nabla_{\bm r} U_{\rm eff}({\bm r},t) = e \nabla_{\bm r} U({\bm r},t)$ and ${\bm E}({\bm r},t) = \nabla_{\bm r} U({\bm r},t)$ is the electric field in the graphene sheet.

The relation between the electric potential and the electron density is found by solving the Poisson equation $\nabla_{\bm r'} \cdot \lbrack \epsilon(z) {\bm E}({\bm r},t) \rbrack = -4 \pi e n({\bm r},t) \delta(z)$.
The average density is $\bar{n} = C U_{0} / e$, where $C = \epsilon_{1} / (4 \pi d)$ is the geometrical capacitance per unit area.
Electric potential and density fluctuations are related by
\begin{equation}\label{eq:channelpotential}
\varphi({\bm k},t)\vert_{z = 0} = - \frac{2 \pi e}{k}\frac{1 - \exp{(-2 d k)}}{\displaystyle \frac{\epsilon_{1} + \epsilon_{2}}{2} - \frac{\epsilon_{2} - \epsilon_{1}}{2} \exp(-2 d k)}n({\bm k},t)~,
\end{equation}
where ${\bm k}$ is a 2D wave vector in the graphene sheet and $\varphi({\bm k},t)\vert_{z = 0}$ and $n({\bm k},t)$ are the 2D Fourier transforms of the electric potential and of the electron density, respectively.
In the limit $d k \to 0$, in which the distance between the graphene sheet and gate is much smaller than the typical wavelength of density oscillations, the relation (\ref{eq:channelpotential}) between the density and gate-to-channel swing assumes the local form
\begin{equation}\label{eq:GCA}
n({\bm r},t) = \frac{C}{e} U({\bm r},t)~.
\end{equation}
The latter expression is known as the ``gradual channel'' approximation~\cite{dyakonov_prl_1993, dyakonov_prb_1995, dyakonov_ieee_1996a, dyakonov_ieee_1996b}.
In this regime, the long-range tail of the Coulomb interaction is completely screened by the charges on the gate.
Using the gradual channel approximation, the Euler and continuity equations read
\begin{equation}\label{eq:eulerGCA}
\begin{split}
\partial_{t} {\bm v}({\bm r},t) & + \lbrack {\bm v}({\bm r},t) \cdot \nabla_{\bm r} \rbrack {\bm v}({\bm r},t) = - \frac{v_{\rm F}^{2} e}{\varepsilon_{\rm F}({\bm r},t)} {\bm E}({\bm r},t) \\
& - \frac{v_{\rm F}^{2}}{2 U({\bm r},t)} \nabla_{\bm r} U({\bm r},t)
- \frac{{\bm v}({\bm r},t)}{2 U({\bm r},t)} \partial_{t} U({\bm r},t)
\end{split}
\end{equation}
and
\begin{equation}\label{eq:continuityGCA}
\partial_{t}U({\bm r},t) + \nabla_{\bm r} \cdot \lbrack U({\bm r},t) {\bm v}({\bm r},t) \rbrack  = 0~.
\end{equation}
\subsection{Making contact with the hydrodynamics of 2DEGs}
\label{ssec:2DEGlike}

In this Section we make a crude approximation on the Euler equation (\ref{eq:eulerGCA}) for a 2D MDF fluid to recover earlier results pertaining to the 2DEG literature.

When the two terms on the second line of Eq.~(\ref{eq:eulerGCA}) are neglected and the Fermi energy in the denominator of the first term on the right-hand side of Eq.~(\ref{eq:eulerGCA}) is assumed to be constant in space, $\varepsilon_{\rm F}({\bm r},t) \to \hbar v_{\rm F} k_{\rm F}$, where $k_{\rm F} = \sqrt{4 \pi \bar{n}/N_{\rm f}}$ is the Fermi wave number corresponding to the average electron density $\bar{n}$, Eq.~(\ref{eq:Euler_complete}) reduces to
\begin{equation}\label{eq:euler2DEG}
\partial_{t} {\bm v}({\bm r},t) + \lbrack {\bm v}({\bm r},t) \cdot \nabla_{\bm r} \rbrack {\bm v}({\bm r},t) = - \frac{e}{m_{\rm c}} \nabla_{\bm r} U({\bm r},t)~,
\end{equation}
which is identical to the Euler equation used by DS~\cite{dyakonov_prl_1993, dyakonov_prb_1995, dyakonov_ieee_1996a, dyakonov_ieee_1996b} with the bare electron mass $m$ replaced by the density-dependent cyclotron mass~\cite{geim_naturemater_2007,castroneto_rmp_2009} $m_{\rm c} = \hbar k_{\rm F}/v_{\rm F}$.
The validity of this 2DEG-like approximation to describe the plasma-wave photoresponse of a graphene FET will be critically examined below.

In the following, we assume that our system is translationally invariant in one direction (the one perpendicular to the source-drain direction) and therefore specify Eqs.~(\ref{eq:eulerGCA})-(\ref{eq:continuityGCA}) to a one-dimensional (1D) geometry in the $x$ direction, with $0\leq x \leq L$.

Linearizing Eqs.~(\ref{eq:continuityGCA})-(\ref{eq:euler2DEG}) we obtain
\begin{equation}
\partial_{t}U(x,t) + U_{0} \partial_{x}v(x,t)  = 0
\end{equation}
and
\begin{equation}
\partial_{t}v(x,t) + \frac{e}{m_{\rm c}} \partial_{x} U(x,t)  = 0~.
\end{equation}
It is well known that these two coupled equations admit solutions in the form of traveling waves
\begin{equation}
v(x,t), \, U(x,t) \propto e^{-i \omega t} e^{i k x} + \mbox{c.c.}~,
\end{equation}
with $\omega = s k$.
The plasma-wave velocity is given by~\cite{principi_ssc_2011,svintsov_jap_2012}
\begin{equation}\label{eq:plasmaspeed}
s \equiv \sqrt{\frac{e U_{\rm 0}}{m_{\rm c}}} = v_{\rm F} \sqrt{N_{\rm f} \alpha_{\rm ee} d k_{\rm F}}~,
\end{equation}
where 
\begin{equation}\label{eq:alphaeegate}
\alpha_{\rm ee} \equiv \frac{e^{2}}{\hbar v_{\rm F} \epsilon_{1}}~
\end{equation}
is a dimensionless coupling constant that controls the strength of e-e interactions~\cite{footnote1}.
As shown above, for typical carrier densities and device sizes, $s \gg v_{\rm F}$.
Our derivation, however, does not formally rely on this inequality and applies also to the regime in which the Fermi velocity and the plasma wave  velocity are comparable, a situation which is achieved when $d k_{\rm F} \sim 1$.

\subsection{Expansion in powers of density fluctuations}

The Euler equation (\ref{eq:eulerGCA}) is not amenable to analytic treatment.
We therefore introduce an expansion of this equation in powers of the deviation $\delta n(x,t) \equiv n(x,t) -{\bar n}$ of the local density $n(x,t)$ from its average value $\bar{n}$.
Because of the gradual channel approximation (\ref{eq:GCA}), a similar expansion can be carried out in powers of $\delta U(x,t) \equiv U(x,t) - U_0$.  

At this stage, it is also convenient to introduce {\it dimensionless} quantities by scaling lengths, electrical potentials, velocities, and time with $L$, $U_{0}$, $s$, and $L/s$, respectively.
For the sake of notational simplicity, dimensionless variables will be denoted with the same symbols used for dimensionful variables.

Expanding the Euler equation (\ref{eq:eulerGCA}) up to {\it second order} in $\delta U(x,t)$ we obtain
\begin{equation}\label{eq:eulerMDF}
\begin{split}
\partial_{t} v(x,t) & + v(x,t) \partial_{x} v(x,t) = - \partial_{x} \delta U(x,t)  \\
& + \frac{1}{2} \delta U(x,t)  \partial_{x} \delta U(x,t)\\
& - \frac{1}{2} v_{\rm F}^{2} \left \lbrack  \partial_{x} + \frac{v(x,t)}{v_{\rm F}^{2}} \partial_{t} \right  \rbrack \delta U(x,t)  \\
& + \frac{1}{2} v_{\rm F}^{2} \delta U(x,t)  \left \lbrack  \partial_{x} + \frac{v(x,t)}{v_{\rm F}^{2}} \partial_{t} \right  \rbrack  \delta U(x,t) ~.
\end{split}
\end{equation}
Once again, all the quantities appearing in the previous equation ($x$, $t$, $\delta U$, etc.) are dimensionless.
The first two terms in the right-hand side of Eq.~(\ref{eq:eulerMDF}) stem from the expansion in powers of $\delta U(x,t)$ of the term in Eq.~(\ref{eq:eulerGCA}) which is proportional to ${\bm E}({\bm r},t)/\varepsilon_{\rm F}({\bm r},t) = \nabla_{\bm r}U({\bm r},t)/\varepsilon_{\rm F}({\bm r},t)$.
The third and the fourth term originate from $\nabla_{\bm r} P({\bm r},t)$ and $\partial_t P({\bm r},t)$.
The first line in Eq.~(\ref{eq:eulerMDF}) reproduces the 1D version of Eq.~(\ref{eq:euler2DEG}).
The terms contained in the last three lines of Eq.~(\ref{eq:eulerMDF}) are qualitatively new terms pertaining to the hydrodynamic theory of the 2D MDF fluid.
They do not have an equivalent in a 2DEG.
The implications of these nonlinear terms on the plasma-wave photoresponse of a gated graphene sheet will be the subject of the reminder of this Article.

We now proceed to estimate the relative magnitude of each term in the right-hand side of Eq.~(\ref{eq:eulerMDF}).
We assume a profile $\delta U(x,t)  = U(k,\omega)\cos(k x - \omega t)$ with $\omega = k$.
In this case, the differential operators simplify to $\partial_{x} \to k$ and $\partial_{t} \to \omega$, and $k$ can be collected as a common prefactor.
We remind the reader that $v(x,t) \ll v_{\rm F}$.
As discussed above, typically $v_{\rm F} \ll 1$ (the Fermi velocity is much smaller than the plasma-wave group velocity when $d k_{\rm F}\gg 1$).
The first nonlinear term on the right-hand side of Eq.~(\ref{eq:eulerMDF}) can be estimated as 
\begin{equation}\label{eq:fluctuationsmass}
\frac{1}{2} \delta U(x,t)  \partial_{x} \delta U(x,t) \approx \frac{k}{2}[U(k,\omega)]^2~.
\end{equation}
Similarly,
\begin{equation}
\begin{split}
\frac{1}{2} v_{\rm F}^{2} \left\lbrack \partial_{x} \vphantom{\frac{v(x,t)}{v_{\rm F}^{2}}} \right. & + \left.\frac{v(x,t)}{v_{\rm F}^{2}} \partial_{t} \right\rbrack \delta U(x,t) \\
& \approx \frac{k}{2}\left[v^2_{\rm F} + v(x,t)\right]U(k,\omega)~.
\end{split}
\end{equation}
We therefore see that the term (\ref{eq:fluctuationsmass}) does not contain velocity factors.
Keeping only this term in Eq.~(\ref{eq:eulerMDF}) can be seen as a ``zeroth-order expansion'' in powers of both $v(x,t) \ll 1$ and $v_{\rm F} \ll 1$.

\section{Generation of plasma waves}
\label{sect:instability}
\subsection{Plasma waves in a 2D MDF fluid moving with a constant speed}
\label{ssec:linearization}

In this Section we study the dispersion of plasma waves in a 2D MDF fluid that moves with an average constant speed $v_0$, measured in units of the plasma-wave speed $s$ defined in Eq.~(\ref{eq:plasmaspeed}).  
We repeat the analysis of Sect.~\ref{ssec:2DEGlike}, allowing, however, for $v_0 \neq 0$.

To this aim, we discard all the terms in Eq.~(\ref{eq:eulerMDF}) which are ${\cal O}(\delta U^2)$ obtaining
\begin{equation}\label{eq:eulerLin}
\begin{split}
\partial_{t} v(x,t) & + v(x,t) \partial_{x} v(x,t) = - \partial_{x} \delta U(x,t) \\
& - \frac{1}{2} v_{\rm F}^{2} \left \lbrack  \partial_{x} + \frac{v(x,t)}{v_{\rm F}^{2}} \partial_{t} \right  \rbrack  \delta U(x,t)~.
\end{split}
\end{equation}
As usual, the previous equation must be supplemented by the continuity equation.
We now seek solutions of the form $\delta U(x,t) = U_1(x,t)$ and $v(x,t) = v_0 + v_1(x,t)$, with
\begin{equation}\label{eq:travelingwaves}
v_1(x,t), \, U_1(x,t) \propto e^{-i \omega t} e^{i k x} + \mbox{c.c.}~.
\end{equation}
We find that for each value of $\omega$, two values of $k$ are possible, corresponding to waves traveling in opposite directions.
In the limit $v_0 \ll v_{\rm F}$, the dispersion reads
\begin{equation}\label{eq:relvelsum}
\omega \simeq \left \lbrack \frac{3}{4}v_{0} \pm s(v_{\rm F}) \right \rbrack k~,
\end{equation}
where
\begin{equation}\label{eq:relsoundspeed}
s(v_{\rm F}) \equiv \sqrt{1 + v_{\rm F}^{2}/2}~.
\end{equation}
The previous result differs from Eq.~(\ref{eq:plasmaspeed}) because we have taken into account the contribution of the pressure to the restoring force, i.e.~the term $-v^2_{\rm F}\partial_x \delta U(x,t)/2$ in Eq.~(\ref{eq:eulerLin}).
The correction $v_{\rm F}^{2} / 2$ in Eq.~(\ref{eq:relsoundspeed}) can therefore be understood from the following general argument.
From Eqs.~(\ref{eq:fermienergy}) and (\ref{eq:pressure}) we see that the 2D MDF fluid fulfills a {\it polytropic} equation of state, $P = C \rho^{\gamma}$, $\rho = m_{\rm c} {\bar n}$ being the mass density and $\gamma = 3/2$.
Then, the usual thermodynamic expression for the sound speed gives $s_{\rm P} = \gamma P /\rho = v_{\rm F}/\sqrt{2}$.

Before concluding, we would like to comment on the factor $3/4$ which appears in front of $v_0$ in Eq.~(\ref{eq:relvelsum}).
From Eq.~(\ref{eq:relvelsum}) with $v_{0} = 0$, we see that $s(v_{\rm F})$ is the plasma-wave speed in a reference system which moves with velocity $v_{0}$ with respect to the laboratory rest frame.
Contrary to the 2DEG case, however, the plasma-wave speed in the laboratory rest frame cannot simply be obtained by performing a Galilean transformation $v \to v_{0} + v$.
In the case of present interest, we find that there is a correction of magnitude $v_0/4$ which reduces (increases) the plasma-wave speed, with respect to the result of the Galilean transformation $v_{0} \pm s(v_{\rm F})$.
This fact is well-known in the context of relativistic hydrodynamics~\cite{font_aa_1994}.

For future purposes, we introduce the parameter
\begin{equation}\label{eq:xi}
\xi \equiv \frac{v_{\rm F}}{s(v_{\rm F})} = \frac{1}{\sqrt{N_{\rm f}\alpha_{\rm ee} k_{\rm F} d +1/2}}~.
\end{equation}

\subsection{Plasma-wave instability}

The dispersion relation (\ref{eq:relvelsum}) characterizes the collective modes of a 2D MDF fluid in the presence of a metal gate and in the thermodynamic limit.

DS showed~\cite{dyakonov_prl_1993} that appropriate boundary conditions in a device of length $L$ define a resonator which supports discrete modes with frequencies $\lbrace \omega_{n} \rbrace_{n \in {\mathbb N}}$.
Moreover, when the average drift velocity $v_{0}$ varies in a certain interval, the resonating modes become unstable and their amplitude increases exponentially with a rate $\Gamma = \Im m \lbrack \omega_{n} \rbrack$, which is independent of $n$. 

The DS boundary conditions for the instability are
\begin{equation}\label{eq:bcinst}
U(x = 0,t) = 1, \quad j(x=1,t) = v_{0}~,
\end{equation}
where $j(x,t) = U(x,t)v(x,t)$ is the current density.
The boundary conditions correspond to a constant density at the source ($x = 0$) and a constant current at the drain ($x = 1$).
We study the DS instability with both space and time derivatives of the pressure taken into account.
However, we neglect momentum relaxation, which takes place on a time scale $\tau$.
This approximation is consistent if we assume that the exponential growth of the amplitude of the modes is quenched by nonlinear terms in the equations of motion on a time scale shorter than $\tau$.

The boundary conditions for the fluctuations, which follow from Eq.~(\ref{eq:bcinst}), read
\begin{equation}\label{eq:bclininst}
U_{1}(x = 0,t) = 0, \quad v_{1}(x=1,t) + v_{0} U_{1}(x=1,t) = 0~.
\end{equation}
To find the allowed values of $\omega$, we insert the ansatz~(\ref{eq:travelingwaves}) into Eq.~(\ref{eq:bclininst}) and solve for $\omega$ to linear order in $v_{0}$.
We find that the real part of the frequency is of the form
 \begin{equation}\label{spagnasuca}
\Re e \lbrack \omega_{n} \rbrack = s(v_{\rm F}) \frac{\pi}{2} n~, 
\end{equation}
where $n$ is a positive integer, while the imaginary part is 
$\Gamma = \Im m \lbrack \omega_{n} \rbrack = 3 v_{0} / 4$.
The fundamental $n=1$ plasma resonance defines the Fermi-velocity-dependent plasma frequency
\begin{equation}\label{eq:omegaPvF}
\omega_{\rm P}(v_{\rm F}) \equiv \frac{\pi}{2} \sqrt{1 + \frac{v_{\rm F}^{2}}{2}}~.
\end{equation}
With respect to the ordinary 2DEG~\cite{dyakonov_prl_1993}, the instability rate $\Gamma$ is reduced by a factor $3/4$. 
As noted above, this effect is of purely relativistic origin and stems from the non-Galilean nature of electron dynamics in graphene.

\section{Resonant plasma-wave photoresponse}
\label{sect:photoresponse}
\begin{figure}
\begin{overpic}[width=\linewidth]{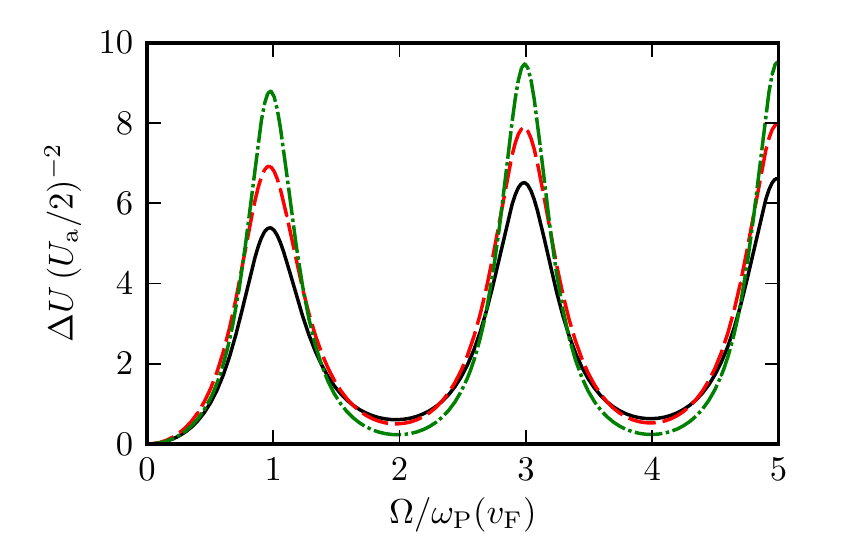}\put(2,60){}\end{overpic}
\caption{\label{fig:photovoltage}
(Color online) The dc photovoltage $\Delta U$ (measured in units of the incident power) as a function of the frequency $\Omega$ of the incoming radiation [measured in units of the plasma-wave frequency (\ref{eq:omegaPvF})]. The solid line represents the dc photovoltage calculated within the 2DEG-like approximation discussed in Sect.~\ref{ssec:2DEGlike} [$C_{1}$,$C_{2}\to 0$ in Eq.~(\ref{eq:photovoltage})]. The dashed line is the photovoltage obtained by including the contribution of the local fluctuations of the cyclotron mass [$C_{1}\to 0$ and $C_2 \neq 0$ in Eq.~(\ref{eq:photovoltage})]. Finally, the dash-dotted line is the full expression in Eq.~(\ref{eq:photovoltage}) ($C_1, C_2 \neq 0$).
In the last two cases, $v_{\rm F}$ is obtained from Eq.~(\ref{eq:plasmaspeed}) with $\alpha_{\rm ee} \times d k_{\rm F} = 0.18$.
We remind the reader that the ratio reported on the horizontal axis depends on the magnitude of $v_{\rm F}$.
In all three cases, $\tau = 1.0~L/s$. }
\end{figure}

We now consider the response of the system to a periodic modulation of the gate-to-channel swing at the source, with vanishing current at the drain.
The periodic modulation at the source can be induced by an impinging radiation (see Fig.~\ref{fig:one}), collected, for instance, by means of an appropriate antenna~\cite{vicarelli_naturemat_2012}.

We want to solve Eq.~(\ref{eq:eulerMDF}) together with the continuity equation
\begin{equation}\label{eq:continuity1D}
\partial_{t}U(x,t) + \partial_x\lbrack U(x,t)v(x,t)\rbrack  = 0~,
\end{equation}
with the DS boundary conditions for detection:
\begin{equation}\label{eq:bcphotoresp}
U(x = 0,t) = 1 + U_{\rm a}\cos(\Omega t), \quad j(x=1,t) = 0~.
\end{equation}
Here $\Omega$ is the frequency of the incoming radiation.
Our aim is to demonstrate that a dc potential difference $\Delta U$ is generated between the drain and source in response to the oscillating radiation and {\it in the absence of a source-drain bias}. 
Most importantly, we will elucidate the contributions to $\Delta U$ arising from the peculiar relativistic corrections to the Euler equation of motion pertaining to graphene.

Following DS~\cite{dyakonov_ieee_1996a}, we seek solutions of the Euler (\ref{eq:eulerMDF}) and continuity (\ref{eq:continuity1D}) equations in the form of power expansions:
\begin{equation}\label{eq:expansionv}
v(x,t) = \epsilon v_{1}(x,t) + \epsilon^{2}\lbrack \delta v(x) + v_{2}(x,t) \rbrack + \dots
\end{equation}
and
\begin{equation}\label{eq:expansionU}
U(x,t) = 1 + \epsilon U_{1}(x,t) + \epsilon^{2}\lbrack \delta U(x) + U_{2}(x,t) \rbrack + \dots~.
\end{equation}
Here, $\epsilon$ is a dimensionless parameter that helps with the bookkeeping while solving Eqs.~(\ref{eq:eulerMDF}) and~(\ref{eq:continuity1D}) perturbatively.
The functions $v_1(x,t)$, $U_1(x,t)$ and $v_2(x,t)$, $U_2(x,t)$ are assumed to oscillate in time with frequency $\Omega$ and $2\Omega$, respectively, while $\delta v(x)$ and $\delta U(x)$ are constant in time.
With the ansatz (\ref{eq:expansionU}) the dc photovoltage is $\Delta U \equiv \delta U(1) - \delta U(0)$.
This perturbative approach is fully justified when the external perturbation is sufficiently weak, i.e.~when $U_{\rm a} = {\cal O}(\epsilon)$.
We find that $\Delta U  = {\cal O}(\epsilon^2)$, i.e.~the dc photovoltage is proportional to the power of the incoming radiation.

To linear order, the equations of motion read:
\begin{equation}\label{eq:eulerlin}
\partial_{t} v_{1}(x,t) = - (1 + v_{\rm F}^{2}/2) \partial_{x} U_{1}(x,t) - v_1(x,t) / \tau
\end{equation}
and
\begin{equation} \label{eq:contlin}
\partial_{t} U_{1}(x,t) + \partial_{x} v_{1}(x,t) = 0~, 
\end{equation}
with boundary conditions
\begin{equation}\label{eq:bclinphotoresp}
U_{1}(x=0,t) = U_{\rm a} \cos{(\Omega t)},\quad v_{1}(x=1,t) = 0~.
\end{equation}

Note that in Eq.~(\ref{eq:eulerlin}) we have included a linear phenomenological friction term, proportional to the inverse momentum relaxation time $\tau^{-1}$. 

The solution of the linear system of differential equations (\ref{eq:eulerlin})-(\ref{eq:contlin}) is given by
\begin{equation}\label{eq:linearSolutionDS}
v_{1}(x,t) = \frac{U_{\rm a}}{2} \frac{\Omega}{K} \left \lbrack  \frac{e^{i K x}}{1+e^{2 i K}} - \frac{e^{-i K x}}{1+e^{-2 i K}} \right  \rbrack  e^{-i \Omega t} + \mbox{c.c.}
\end{equation}
and
\begin{equation}
U_{1}(x,t) = \frac{U_{\rm a}}{2} \left \lbrack  \frac{e^{i K x}}{1+e^{2 i K}} + \frac{e^{-i K x}}{1+e^{-2 i K}} \right  \rbrack  e^{-i \Omega t} + \mbox{c.c.}~,
\end{equation}
with
\begin{equation}\label{eq:dispersionK}
K = K(\Omega) = \frac{\Omega}{s(v_{\rm F})} \sqrt{1 + \frac{i}{\Omega \tau}}~.
\end{equation}
Note that in the limit $\Omega \tau \gg 1$ Eq.~(\ref{eq:dispersionK}) coincides with Eq.~(\ref{eq:relvelsum}) with $v_{0} = 0$.

\begin{figure}
\begin{overpic}[width=\linewidth]{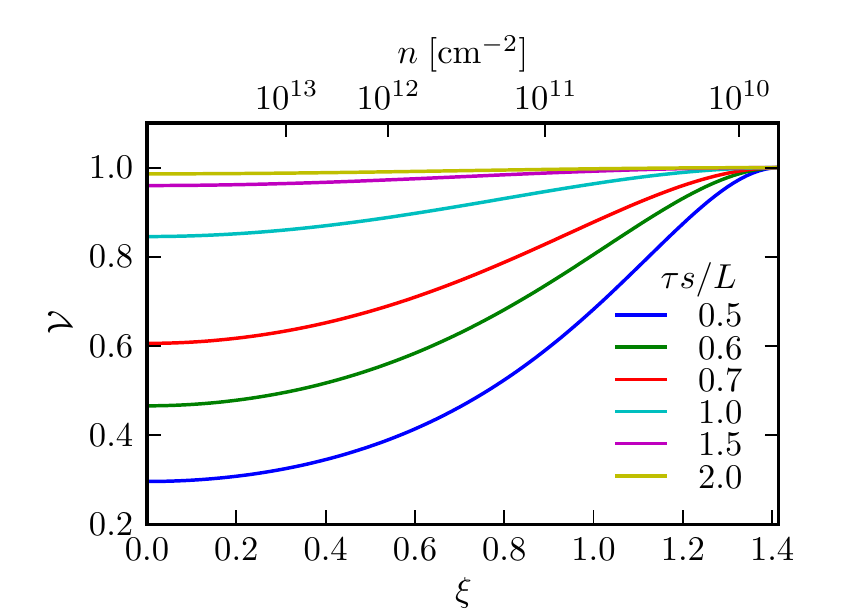}\put(2,60){(a)}\end{overpic}
\begin{overpic}[width=\linewidth]{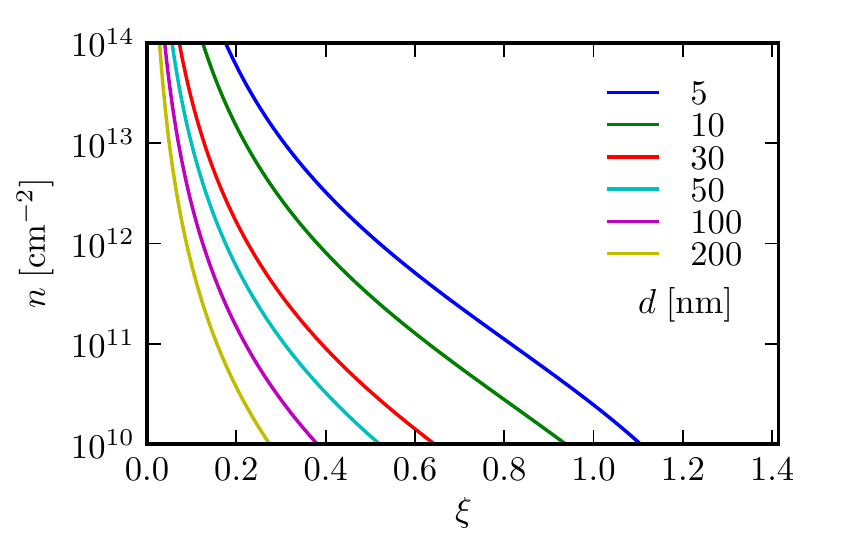}\put(2,60){(b)}\end{overpic}
\caption{\label{fig:visibility}
(Color online) (a) The visibility ${\cal V}$ of the photoresponse as a function of $\xi$ as defined in Eq.~(\ref{eq:xi}).
Different curves correspond to different values of the momentum relaxation time $\tau$ measured in units of the channel length $L$ divided by the plasma wave speed $s$: from bottom to top the value of $\tau$ increases.
The top horizontal axis reports the density ${\bar n}$ corresponding to the value of 
$\xi$ reported in the bottom horizontal axis: in this case we have fixed $d = 10~{\rm nm}$ and $\alpha_{\rm ee} = 0.9$.
(b) The average carrier density ${\bar n}$ (in units of $10^{12}~{\rm cm}^{-2}$) as a function of $\xi$.
Different curves correspond to different values of $d$ (in units of ${\rm nm}$): from top to bottom $d$ increases. }
\end{figure}

We then write the Euler equation to second order in $\epsilon$ and average it over one period of the incoming radiation.
We find 
\begin{equation}\label{eq:euleraverage}\begin{split}
\langle v_{1}(x,t) \partial_{x}v_{1}(x,t) \rangle_{t} & = - (1 + v_{\rm F}^{2}/2) \partial_{x} \delta U(x) - \frac{1}{\tau} \delta v(x) \\
& + \frac{1}{2}(1 + v_{\rm F}^{2}) \langle U_{1}(x,t) \partial_{x} U_{1}(x,t) \rangle_{t} \\
& - \frac{1}{2} \langle v_{1}(x,t) \partial_{t} U_{1}(x,t) \rangle_{t}~,
\end{split}\end{equation}
where $\langle f(t) \rangle_{t} \equiv T^{-1} \int_{0}^{T} dt f(t)$ denotes the time average over one period $T = 2\pi/\Omega$ of the external radiation.

Eq.~(\ref{eq:euleraverage}) depends on $\delta v(x)$---see Eq.~(\ref{eq:expansionv}).
This quantity can be easily obtained by averaging over time the continuity equation, written up to second order in $\epsilon$.
We find 
$\delta v(x) = - \langle v_{1}(x,t) U_{1}(x,t) \rangle_{t}$.
The final expression for the photovoltage is
\begin{equation}\begin{split}
\Delta U  = & \frac{1}{1 + v_{\rm F}^{2} / 2} \left \lbrace \frac{1}{2} \langle v_{1}(0,t)^{2} \rangle_{t} \right . \\
& + \frac{1}{4}(1 + v_{\rm F}^{2}) \lbrack \langle U_{1}(1,t)^{2} \rangle_{t} - \langle U_{1}(0,t)^{2} \rangle_{t} \rbrack \\
& + \frac{1}{\tau} \int_{0}^{1} dx \langle v_{1}(x,t) U_{1}(x,t) \rangle_{t} \\
& \left . - \frac{1}{2} \int_{0}^{1} dx \langle v_{1}(x,t) \partial_{t} U_{1}(x,t) \rangle_{t} \vphantom{\frac{1}{2}}  \right \rbrace~.
\end{split}\end{equation}
Remarkably, we have found an analytical expression for $\Delta U$ which can be given in the following rather compact form:
\begin{equation}\label{eq:photovoltage}
\begin{split}
\Delta U & = \left ( \frac{U_{\rm a}}{2} \right )^{2} \left \lbrace 1 - \frac{C_{2}}{2} - \frac{2 - C_{2}}{\cos{2 K_{1}} + \cosh{2 K_{2}}} \right . \\
& + \left . \beta (1 - C_{1}) \frac{\cosh{2 K_{2}} - \cos{2 K_{1}}}{\cos{2 K_{1}} + \cosh{2 K_{2}}} \right \rbrace~,
\end{split}\end{equation}
where $C_{1} = 1/4$, $C_{2} = (1 + v_{\rm F}^{2}) / (1 + v_{\rm F}^{2} / 2)$, $\beta = 2 \Omega \tau / \sqrt{1 + (\Omega \tau)^{2}}$, and $K_{1}$ ($K_{2}$) is the real (imaginary) part of $K$, which reads
\begin{equation}
K_{1,2} = \frac{\Omega}{s(v_{\rm F})} \frac{1}{\sqrt{2}} \sqrt{ \sqrt{1 + \frac{1}{(\Omega \tau)^{2}}} \pm 1}~.
\end{equation}
Eq.~(\ref{eq:photovoltage}) reduces to the DS result for an ordinary 2DEG~\cite{dyakonov_ieee_1996a} (with the cyclotron mass $m_{\rm c}$ playing the role of the electron mass $m$) in the limit $C_1 = C_2 =0$.
The terms controlled by $C_2$ stem from the local fluctuations of the cyclotron mass.
This coefficient tends to unity in the limit $v_{\rm F} \ll 1$.
The terms controlled by $C_1$ originate from the derivatives of the pressure and suppress the photoresponse with respect to the case $C_1 =0$. 

Illustrative plots of $\Delta U$ as a function of the incoming radiation frequency $\Omega$ are presented in Fig.~\ref{fig:photovoltage} for a given value of the momentum relaxation time $\tau$.
The photoresponse features maxima at odd multiples of the plasma frequency (\ref{eq:omegaPvF}).
Nervous readers can easily convert dimensionless values of the momentum relaxation time $\tau$ into values of the mobility $\mu$.
The mobility $\mu$ is obtained from Eq.~(\ref{eq:eulerlin}) in the steady-state regime (i.e.~when the time derivative on the left-hand side is neglected) and reads $\mu = (1 + v_{\rm F}^{2}/2) \tau$.
Restoring, for a moment, physical units, the previous relation reads:
\begin{equation}
\mu =  \frac{e \tau}{m_{\rm c}} \left(1 + \frac{v_{\rm F}^{2}}{2s^2} \right) = \frac{e}{\hbar} \frac{v_{\rm F} \tau}{\pi n d} \left (d\sqrt{\pi n} + \frac{1}{8 \alpha_{\rm ee}} \right )~.
\end{equation}
More conveniently,
\begin{equation}
\mu = 5.0 \times 10^{5} \frac{\tau}{nd} \left ( 0.18 d \sqrt{n} + \frac{1}{8 \alpha_{\rm ee}} \right )~{\rm cm}^2/({\rm V} {\rm s})~,
\end{equation}
where $\tau$, $n$, and $d$ must be expressed in ${\rm ps}$, $10^{12}~{\rm cm}^{-2}$, and ${\rm nm}$, respectively.

The photovoltage (\ref{eq:photovoltage}) depends sensibly on the ratio $\xi$ between the Fermi velocity $v_{\rm F}$ and the effective plasma-wave speed $s(v_{\rm F})$. 
From Eq.~(\ref{eq:xi}) we see that this ratio can be tuned in the range $[0,\sqrt{2}]$ by changing the distance $d$ between graphene and the gate, the dielectric constant $\epsilon_1$, and the average carrier density ${\bar n}$.
Fig.~\ref{fig:photovoltage} shows that the photoresponse of a system that is described by relativistic hydrodynamics can be tuned to regimes where its resonant maxima have larger amplitude and its minima are shallower.
We are therefore naturally led to introduce the visibility ${\cal V}$ of the resonant photovoltage profile according to
\begin{equation}\label{eq:visibility}
{\cal V} \equiv \frac{\Delta U_{\rm max} - \Delta U_{\rm min}}{\Delta U_{\rm max} + \Delta U_{\rm min}}~,
\end{equation}
where $\Delta U_{\rm max}$ ($\Delta U_{\rm min}$) is the maximum (minimum) of the photoresponse evaluated at $\Omega \approx \Re e[\omega_1]$ ($\Omega \approx \Re e[\omega_2]$)---see Eq.~(\ref{spagnasuca}).
Illustrative plots of ${\cal V}$ are shown in Fig.~\ref{fig:visibility}.
We see that, for all values of the momentum relaxation time $\tau$, the visibility drastically increases as $\xi$ increases.
In particular, in the limit $\xi \to \sqrt{2}$ the visibility reaches its maximum value. Reaching values of $\xi$ of the order of $\sqrt{2}$ requires to minimize the product $dk_{\rm F}$. Values of this product as small as $0.1$ can be reached by placing graphene on high-quality substrates like h-BN, where carrier densities as low as $\sim 10^{10}~{\rm cm}^{-2}$ can be reached without entering the regime where disorder 
(electron-hole puddles) dominates. In these samples, gold metal gates can be placed as close as $10~{\rm nm}$ to the graphene sheet, as recently shown in a quantum capacitance measurement carried out in single-layer graphene~\cite{manchester}.

\section{Conclusions and perspectives}
\label{sect:conclusions}

In this Article we have presented a theory of a photoresponse mechanism which is based on the excitation of plasma waves in the channel of a graphene field-effect transistor (FET).
The FET is coupled to a source of radiation which periodically modulates the voltage difference between the gate and source.
Our main result is a fully analytical expression---Eq.~(\ref{eq:photovoltage})---for the dc voltage difference $\Delta U$ between the drain and source, in response to the oscillating radiation field.

Plasma waves in the channel are described within a hydrodynamic model which takes into account the linear dispersion of massless Dirac fermions in graphene.
Formally, the hydrodynamic theory of two-dimensional massless Dirac fermions in graphene presents strong analogies with the hydrodynamic theory of a relativistic fluid~\cite{Landau06}, with the Fermi velocity $v_{\rm F}$ playing the role of the speed of light $c$.
In particular, we have shown that this yields a breakdown of Galilean invariance whereby the speed of plasma waves in the frame of the moving fluid is not connected by a Galilean transformation to that in the laboratory frame.  
Moreover, the photoresponse substantially depends on the peculiar nonlinearities of the equations of motion which arise due to the formal relativistic nature of the energy-momentum dispersion of carriers in graphene.
We have shown that it is possible to leverage the effect of these nonlinearities to increase the photoresponse of a gated graphene sheet with respect to that occurring in an ordinary two-dimensional electron gas.
Measuring the photoresponse of a graphene sheet in a FET geometry offers the opportunity to demonstrate the importance of relativistic corrections to the ordinary hydrodynamic theory.

Finally, we emphasize that the present theory assumes that the gate of the FET is as long as the channel.
In certain experimental situations~\cite{vicarelli_naturemat_2012} it may be more convenient to exploit gates which are much shorter than the FET channel.
In this case the theory of this Article needs to be changed.
The dependence of the DS-like dc photovoltage on the position of the gate with respect to the source, say, may present non-trivial features.
In the case of a short gate, however, the dc photovoltage will suffer~\cite{vicarelli_naturemat_2012} from contributions of thermoelectric and photoconductive origin.
Last but not least, we would like to emphasize that Vicarelli {\it et al.}~\cite{vicarelli_naturemat_2012} have also shown that {\it bilayer graphene} FETs can be used to fabricate plasma-wave THz photodetectors with low noise-equivalent power.
In the case of bilayer graphene, we note that modulating the gate-to-source voltage will not only modulate the density in the FET channel but also the band gap~\cite{fogler_prb_2010}.
These issues are well beyond the scope of the present Article and will be investigated in forthcoming publications.

\begin{acknowledgments}
It is a pleasure to thank Vittorio Pellegrini and Alessandro Tredicucci for useful discussions.
This work was supported by the Italian Ministry of Education, University, and Research (MIUR) through the program ``FIRB - Futuro in Ricerca 2010'' Grant No.~RBFR10M5BT (``PLASMOGRAPH: plasmons and terahertz devices in graphene'').
We have made use of free software (www.gnu.org, www.python.org).
\end{acknowledgments}

\appendix

\section{Kinetic and hydrodynamic equations}

For the sake of completeness, in this Appendix we summarize  the derivation of the hydrodynamic equations for graphene, starting from the Boltzmann semiclassical equation~\cite{KadanoffBaym,Landau05} for 2D MDFs~\cite{bistritzer_prb_2009,ryzhii_jap_2007a,svintsov_jap_2012}:
\begin{eqnarray}\label{eq:kineticequation}
\lbrack \partial_{t} + {\bm v}_{{\bm k}} \cdot \nabla_{\bm r} &\mp & \frac{1}{\hbar} (\nabla_{\bm r} U_{\rm eff}({\bm r},t) \cdot \nabla_{\bm k})  \rbrack  f_{\rm e/h}({\bm k},{\bm r},t) \nonumber \\ 
&=& {\cal I}[f_{\rm e}, f_{\rm h}]~.
\end{eqnarray}
Here, $f_{\rm e/h}({\bm k},{\bm r}, t)$ represents the probability to find an electron (hole) with momentum $\hbar{\bm k}$, at position ${\bm r}$ and time $t$.
The relation between velocity ${\bm v}_{\bm k}$ and wave vector ${\bm k}$ is
\begin{equation}
{\bm v}_{\bm k} = v_{\rm F} \frac{{\bm k}}{|{\bm k}|}~.
\end{equation}
It is important to note that this relation, peculiar to MDFs, is strongly nonlinear~\cite{rudin_ijhses_2011,mikhailov_njp_2012}.
In the ordinary Schr\"{o}dinger case ${\bm v}_{{\bm k}} = \hbar {\bm k} / m$.
In Eq.~(\ref{eq:kineticequation}) $U_{\rm eff}({\bm r},t)$ is the effective potential energy felt by an electron at position ${\bm r}$ and time $t$, which includes the effect of external potentials. 
The self-consistent electrostatic potential generated by the electron distribution (i.e.~the RPA contribution to the electron self-energy~\cite{KadanoffBaym}) 
should be included in $U_{\rm eff}({\bm r},t)$~\cite{Landau05}.
The term ${\cal I}[f_{\rm e}, f_{\rm h}]$ on the right-hand side of Eq.~(\ref{eq:kineticequation}) represents the Boltzmann collision integral, which takes into account the effect of e-e scattering events on the time-evolution of the distribution function.
In graphene, particular care has to be taken in the calculation of this integral~\cite{brida_naturecommun_2013,tomadin_arxiv_2013} due to the existence of peculiar scattering events (Auger processes) which have a vanishing phase space.

As discussed in Sec.~\ref{sect:model}, the hydrodynamic theory relies on the assumption that the system is in local equilibrium with respect to e-e collisions.
Formally, this means that the distribution functions $f_{\rm e/h}$ fulfill ${\cal I}[f_{\rm e},f_{\rm h}]=0$ at each point in space and time.
Quite generally, since each e-e scattering event conserves the number of particles, total momentum, and total energy, the form of the distribution function which nullifies the collision integral is~\cite{bistritzer_prb_2009, svintsov_jap_2012}
\begin{equation}\label{eq:quasieqansatz}
f_{\rm e/h}({\bm k},{\bm r},t) = \frac{1}{e^{ \beta({\bm r},t) \lbrack \hbar v_{\rm F} k - \hbar {\bm v}({\bm r},t) \cdot {\bm k} \mp \mu({\bm r},t) \rbrack} + 1}~.
\end{equation}

Local macroscopic averages ${\cal O}({\bm r},t)$ of wave-vector dependent quantities ${\cal O}_{\bm k}$ are defined by
\begin{equation}\label{eq:averaging}
{\cal O}({\bm r},t) =  \frac{\sum_{\alpha \in \{{\rm e},{\rm h}\}}\int d{\bm k} f_{\alpha}({\bm k},{\bm r},t) {\cal O}_{\bm k} }{ \sum_{\alpha \in \{{\rm e},{\rm h}\}} \int d{\bm k} f_{\alpha}({\bm k},{\bm r},t)}~,
\end{equation}
where
\begin{equation}
\int d{\bm k} \to \sum_{\sigma \in  \lbrace \uparrow,\downarrow \rbrace } \sum_{\ell \in  \lbrace {\rm K},{\rm K}' \rbrace } \int \frac{d^2{\bm k}}{(2\pi)^{2}}
\end{equation}
indicates integration over momentum space and summation over spin and valley degrees of freedom.
It is easy to see that the parameters ${\bm v}({\bm r},t)$, $1/\lbrack k_{\rm B}\beta({\bm r},t) \rbrack$, and $\mu({\bm r},t)$ introduced in Eq.~(\ref{eq:quasieqansatz}) 
have the physical meaning of local average velocity, temperature, and chemical potential, respectively.

In the case of ordinary Schr\"{o}dinger fermions, the relation between average velocity and wave vector follows by linearity from ${\bm v}_{\bm k} = \hbar {\bm k} / m$ and reads ${\bm v}({\bm r},t) = \hbar {\bm k}({\bm r},t)/m$.
In the 2D MDF case, on the contrary, one finds~\cite{rudin_ijhses_2011}
\begin{equation}\label{eq:momentumrudin}
\hbar {\bm k}({\bm r},t) = \frac{3 P({\bm r},t)}{n({\bm r},t)} \frac{1}{1 - {\bm v}({\bm r},t)^{2} / v_{\rm F}^{2}} \frac{{\bm v}({\bm r},t)}{v_{\rm F}^{2}}~,
\end{equation}
where $n({\bm r},t) = n_{\rm e}({\bm r},t) + n_{\rm h}({\bm r}, t)$ and $P({\bm r},t)$ are the density and pressure, respectively.
The pressure is given by~\cite{rudin_ijhses_2011}
\begin{equation}\label{eq:pressurefull}
P({\bm r},t) = \frac{1}{2 \beta({\bm r},t)} \left \lbrack  \frac{n_{\rm e}({\bm r},t) F_{2}(\zeta)}{F_{1}(\zeta)} + \frac{n_{\rm h}({\bm r},t) F_{2}(-\zeta)}{F_{1}(-\zeta)} \right  \rbrack~,
\end{equation}
where
\begin{equation}
n_{\rm e/h}({\bm r},t) \equiv \int d{\bm k} f_{{\rm e}/{\rm h}}({\bm k},{\bm r},t)
\end{equation}
is the electron (hole) density, $\zeta = \mu({\bm r},t) \beta({\bm r}, t)$, $F_{n}(\zeta) = -\Gamma(n + 1) {\rm Li}_{n + 1}(-e^{\zeta})$, and ${\rm Li}_{n}$ is the polylogarithmic function.
Eq.~(\ref{eq:pressurefull}) is the equation of state of 2D MDFs.

The hydrodynamic equations for density (i.e.~the continuity equation), momentum (i.e.~the Euler equation), and energy can be obtained following the standard procedure~\cite{KadanoffBaym} of multiplying the Boltzmann equation by $1$, $\hbar {\bm k}$, and $\varepsilon_{{\bm k}} = \hbar v_{\rm F} |{\bm k}|$, respectively, and averaging the resulting equations as from Eq.~(\ref{eq:averaging}).

The continuity equation has been reported in Eq.~(\ref{eq:continuity}) and has the same form as in the 2DEG case.

The Euler equation reads
\begin{equation}\label{eq:eulerfullrel}\begin{split}
\partial_{t} {\bm k}({\bm r},t) & + \frac{1}{\hbar} \frac{1}{n({\bm r},t)} \nabla_{\bm r} P({\bm r},t) + \lbrack {\bm v}({\bm r},t) \cdot \nabla_{\bm r} \rbrack {\bm k}({\bm r},t) \\
& + \frac{1}{\hbar} \frac{n_{\rm e}({\bm r},t) - n_{\rm h}({\bm r},t)}{n({\bm r},t)} \nabla_{\bm r} U_{\rm eff}({\bm r},t) = 0~.
\end{split}\end{equation}
We remind the reader that the Euler equation is usually~\cite{Landau06} written in terms of the average velocity ${\bm v}({\bm r}, t)$ only.
In the present case of a 2D MDF fluid, however, rewriting Eq.~(\ref{eq:eulerfullrel}) in terms of ${\bm v}({\bm r},t)$ generates a rather complicated expression, due to the nonlinear relation (\ref{eq:momentumrudin}).
Eq.~(\ref{eq:euler}) represents the final result in the ``non-relativistic'' limit $|{\bm v}({\bm r},t)| \ll v_{\rm F}$. 

Finally, the energy equation reads
\begin{equation}\label{eq:energyfullrel}
\partial_{t} \lbrack n({\bm r},t) \varepsilon({\bm r},t) \rbrack + \hbar v_{\rm F}^{2} \nabla_{\bm r} \cdot \lbrack n({\bm r}, t) {\bm k}({\bm r},t) \rbrack - {\cal F} = 0~,
\end{equation}
with
\begin{equation}
{\cal F} = v_{\rm F} \sum_{i=1}^{2} \frac{\partial}{\partial r_{i}} U_{\rm eff}({\bm r},t) \int d{\bm k} |{\bm k}| \frac{\partial}{\partial p_{i}} \lbrack f_{\rm e}({\bm r},t) - f_{\rm h}({\bm r},t) \rbrack~,
\end{equation}
where $r_{i}$ ($p_{i}$) is the $i$th component of ${\bm r}$ (${\bm p}$).
In this Article we have not made use of the energy equation.
We emphasize that a proper description of thermal transport and thermoelectricity requires to take into consideration energy dissipation (e.g.~to lattice phonons) and heat sources.

\end{document}